\documentclass[acmsmall,screen,nonacm]{acmart}

\usepackage{fancyhdr}
\usepackage{fancyvrb}

\VerbatimFootnotes

\AtBeginDocument{
  }

\acmSubmissionID{68}

\begin{document}

\title{The Calysto Scheme Project}

\author{Douglas S. Blank}
\affiliation{
  \institution{Comet ML, Inc.}
  \city{New York}
  \country{USA}
}
\email{doug.blank@gmail.com}

\author{James B. Marshall}
\affiliation{
  \institution{Sarah Lawrence College}
  \streetaddress{1 Mead Way}
  \city{Bronxville}
  \state{New York}
  \postcode{10708}
  \country{USA}
}
\email{jmarshall@sarahlawrence.edu}

\renewcommand{\shortauthors}{D. S. Blank and J. B. Marshall}

\begin{abstract}
Calysto Scheme is written in Scheme in Continuation-Passing Style, and
converted through a series of correctness-preserving program transformations
into Python. It has support for standard Scheme functionality, including
\texttt{call/cc}, as well as syntactic extensions, a nondeterministic operator
for automatic backtracking, and many extensions to allow Python
interoperation. Because of its Python foundation, it can take advantage of
modern Python libraries, including those for machine learning and other
pedagogical contexts. Although Calysto Scheme was developed with educational
purposes in mind, it has proven to be generally useful due to its simplicity
and ease of installation. It has been integrated into the Jupyter Notebook
ecosystem and used in the classroom to teach introductory Programming Languages
with some interesting and unique twists.
\end{abstract}





\maketitle

\fancyfoot[L]{\small \emph{ICFP 2023 Workshop on Scheme and Functional Programming, Seattle, WA}}

\section{Introduction}

This paper describes the development of Calysto Scheme, a full-featured
implementation of the Scheme programming language written in Scheme, which can
be automatically transpiled into other languages, such as Python and C\#
\cite{CalystoScheme}.  Although our path through the development of this system
was circuitous, the completed project ended up being far more interesting (and
useful) than we could have imagined.

\begin{figure}[h]
  \centering
  \includegraphics[width=0.40\textwidth]{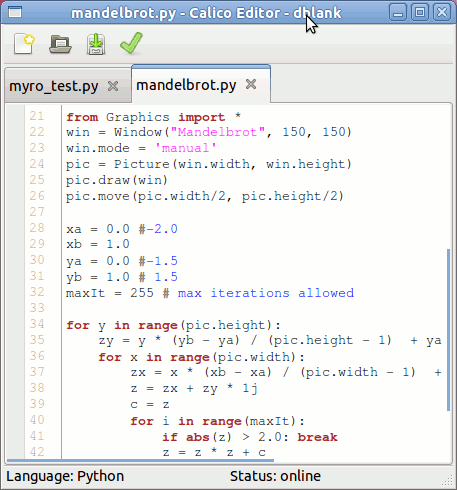}
  \hspace{0.15in}
  \includegraphics[width=0.523\textwidth]{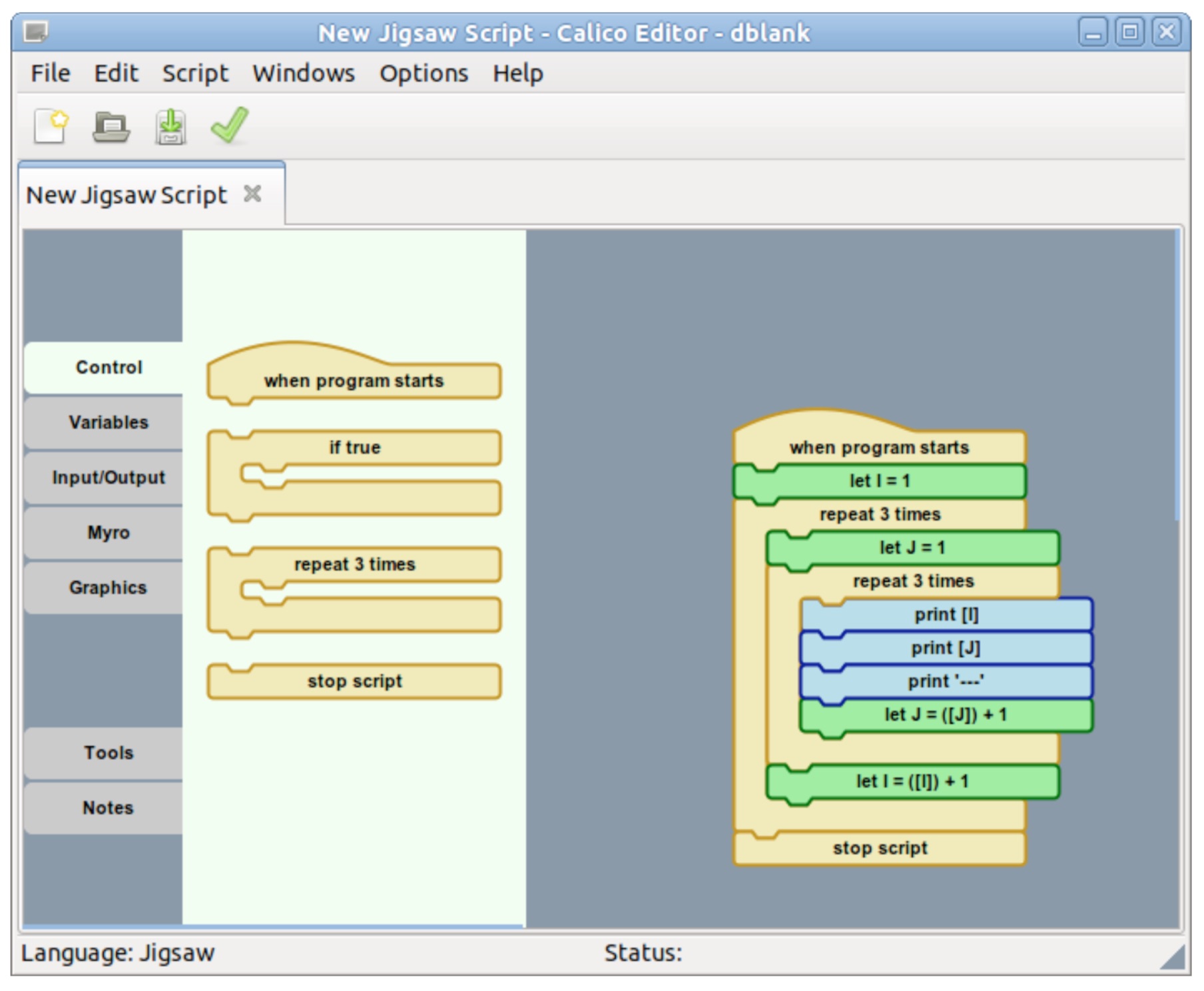}
  \caption{The Calico interface, running Python (left),
    and Jigsaw, a visual block-based language (right).}
  \label{fig:calico}
  \Description{User interface windows showing code written in Python and code
    written in a visual block-based language.}
\end{figure}

The initial motivation for Calysto Scheme grew out of an earlier open-source
project called Calico that was designed to be a multi-programming-language
framework and learning environment for computing education \cite{Calico}.  The
basic premise of this earlier project was to create a common architecture, user
interface, and set of libraries for a variety of programming languages (see
Figure~\ref{fig:calico}).


Many engaging ``pedagogical contexts'' for learning about
computer science have been developed, including media computation
\cite{Guzdial03}, gaming, AI and robotics \cite{Myro}, visualization, music,
and art.  However, these contexts often depend on a set of libraries developed
for a specific programming language, which may constrain the choice of language
if an instructor wishes to have students explore a particular learning
context. Having a common framework separates the details of a specific language
from other pedagogical goals.

One of the interesting aspects of Calico was that instead of having
students learn a different IDE for each language under study, such as
IDLE for Python, or DrRacket for Scheme, students could remain in the
same IDE, but simply switch the programming language. This is similar
in spirit to the way that one can switch languages in DrRacket \cite{Racket}.
However, in Calico, supported languages can easily interoperate through shared
libraries and data structures, even if the languages have little else in common
with each other.

At the time (2007), Microsoft had embarked on a related goal. They were
actively developing what they called the Dynamic Language Runtime (DLR) as part
of the .NET framework. The DLR abstraction layer created a common API and
allowed one to create languages, such as IronPython and IronRuby
\cite{IronRuby}, without having to rewrite the common parts for each
language. Our earlier project adopted the DLR as its foundation, and used these
Iron languages. In addition, the Calico team developed other languages to
augment these, including a visual block-based language.

However, there was a gap in the languages available: there was no
Scheme implementation that could work in this environment. Although
the development of both IronLisp and IronScheme \cite{IronScheme} had
been attempted by other groups, they both ultimately failed for
various reasons, including lack of support for tail-call optimization
(TCO) in the DLR.\footnote{IronScheme eventually did include at
least some support for first-class continuations. However, it was
incrementally added beginning in late 2008
\cite{IronScheme-Continuations} after we had begun building Calysto
Scheme.}

Of course, Scheme can be implemented in languages without TCO. So we set out to
fill this void by developing a Scheme without using the DLR for internal
function calls, but still allowing integration with the DLR for interoperation
with other functionality, such as calling libraries.

But why focus on Scheme? After all, creating a full implementation from scratch
was not a small undertaking, especially since we were not actively involved in
programming languages research. The answer is simple: Scheme is a wonderful
language that captures many sophisticated ideas in a simple syntax. We believed
that students needed to be exposed to these ideas, and we wanted to use a
Scheme that contained all of the features of a modern implementation. We both
studied Computer Science at Indiana University, Bloomington, and were exposed
to the beautiful ideas taught in the Programming Languages courses there, and
felt compelled to do our part in passing the ideas on to the next generation of
students. Thus, we began the development of what would become Calysto Scheme.

\begin{figure}[h]
  \includegraphics[width=0.8\textwidth]{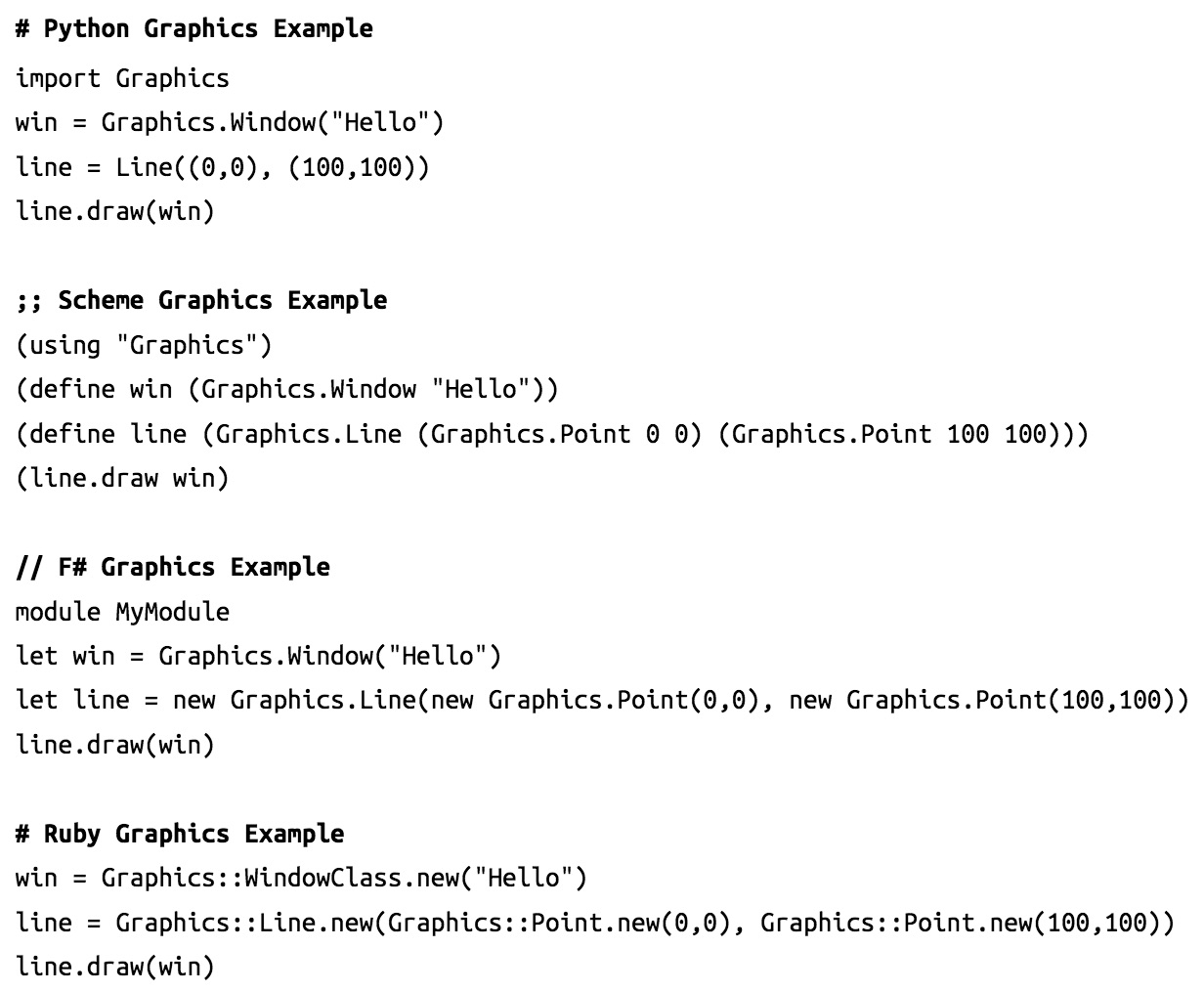}
  \caption{Calling the same graphics library functions from four different languages in Calico.}
  \label{fig:scripts}
  \Description{}
\end{figure}

Once the core Scheme language was written in Scheme, we converted it to a
low-level implementation written in C\#, which could take advantage of the
DLR. The final conversion step to C\# was relatively straightforward. At this
point, we were able to add Scheme to our list of supported languages, including
Python, F\#, and Ruby.  The code snippets in Figure~\ref{fig:scripts} show demo
scripts written in these four languages, all of which call functions from the
same underlying Python graphics library. Each script creates a graphics window
titled ``Hello'', and draws a line between points (0,~0) and (100,~100). The
calls to library functions in each language differ merely in their syntax, but
there are many deeper differences between the languages in terms of semantics.


Although we initially targeted C\# as the implementation language, we
eventually decided to replace C\# with Python (see below). The core Calysto
Scheme architecture remained the same, only the final transformation step
needed to be changed to target Python rather than C\#. The transformation to
Python also included changing calls to C\#'s DLR to calls to standard Python
functions. The following section outlines the overall Calysto Scheme design
pipeline.


\section{Calysto Scheme Design}

Calysto Scheme ensures full support for \texttt{call/cc} and tail-call
optimization, with no limit placed on the depth of the call stack.
Unfortunately, languages such as C\# and Python impose a maximum depth on their
recursion stack and do not support tail-call optimization.  Therefore, our
approach was to implement the Calysto Scheme interpreter in Scheme, and then
automatically convert it, via a series of correctness-preserving program
transformations, into low-level register machine code that does not rely on the
recursion stack, which can then be directly transformed into Python (or C\#) as
the final step.  The high-level Scheme version of the interpreter is written in
Continuation-Passing Style (CPS), with continuations initially represented as
anonymous lambda functions.  The continuations are then converted to a data
structure representation (as lists), and from there passing information to
functions via arguments is replaced by passing information via a set of global
registers, which removes the reliance on the call stack.  At this stage, the
computation is driven by a single ``trampoline'' loop, essentially equivalent
to a while loop \cite{Ganz99, EOPL3}.

As an example of the transformation process, consider a simple recursive
function that adds up the first $n$ positive integers.  We start with a version
of this function written in CPS, using functional continuations, as shown
below.  For example, calling the top-level function \texttt{(sum 100)} returns
5050.

\begin{minipage}{\textwidth}
{\small
\begin{verbatim}
(define sum-cps
  (lambda (n k)
    (if (= n 0)
        (k 0)
        (sum-cps (- n 1)
          (lambda (value)
            (k (+ n value)))))))

;; top-level function
(define sum
  (lambda (n)
    (sum-cps n (lambda (value) value))))

\end{verbatim}
}
\end{minipage}

\noindent
We wrote a program that takes any Scheme program written in CPS, such as the
above, and transforms the code into an equivalent register machine, with
continuations represented as lists.  The resulting register machine for
\texttt{sum} is shown in Figure~\ref{fig:schemeRM}.  Calling \texttt{(sum 100)}
still returns 5050, as before.

At this stage, all functions other than the trampoline simply execute
if-statements or update registers via assignment statements, without ever
calling another function directly (except for low-level built-in primitives
like \texttt{car} or \texttt{+}).  This avoids building up chains of function
calls.  This code can then be converted into Python in a straightforward way
(see Figure~\ref{fig:pythonRM}).  Since the Python version of \texttt{sum} is
no longer constrained by the depth of the recursion stack, it can be called
with arbitrarily large values of $n$.


\begin{figure}[h!]
\begin{minipage}{0.4\textwidth}
{\scriptsize
\begin{verbatim}
;; global registers
(define n_reg 'undefined)
(define k_reg 'undefined)
(define value_reg 'undefined)
(define fields_reg 'undefined)
(define pc 'undefined)
(define final_reg 'undefined)

(define trampoline
  (lambda ()
    (if pc
        (begin
          (pc)
          (trampoline))
        final_reg)))

(define make-cont
  (lambda args
    (cons 'continuation args)))

(define apply-cont
  (lambda ()
    (let ((label (cadr k_reg))
          (fields (cddr k_reg)))
      (set! fields_reg fields)
      (set! pc label))))

(define <cont-1>
  (lambda ()
    (set! final_reg value_reg)
    (set! pc #f)))

(define <cont-2>
  (lambda ()
    (let ((n (car fields_reg))
          (k (cadr fields_reg)))
      (set! k_reg k)
      (set! value_reg (+ n value_reg))
      (set! pc apply-cont))))

(define sum-cps
  (lambda ()
    (if (= n_reg 0)
      (begin
        (set! value_reg 0)
        (set! pc apply-cont))
      (begin
          (set! k_reg (make-cont <cont-2> n_reg k_reg))
        (set! n_reg (- n_reg 1))
        (set! pc sum-cps)))))

;; top-level function
(define sum
  (lambda (n)
    (set! k_reg (make-cont <cont-1>))
    (set! n_reg n)
    (set! pc sum-cps)
    (trampoline)))







\end{verbatim}
}
\caption{Scheme register machine code}
\label{fig:schemeRM}
\end{minipage}
\hspace{0.75in}
\begin{minipage}{0.4\textwidth}
{\scriptsize
\begin{verbatim}
# global registers
n_reg = None
k_reg = None
value_reg = None
fields_reg = None
pc = None
final_reg = None

def trampoline():
    while pc:
        pc()
    return final_reg

def make_cont(*args):
    return ("continuation",) + args

def apply_cont():
    global fields_reg, pc
    label = cadr(k_reg)
    fields = cddr(k_reg)
    fields_reg = fields
    pc = label

def cont_1():
    global final_reg, pc
    final_reg = value_reg
    pc = False

def cont_2():
    global k_reg, value_reg, pc
    n = car(fields_reg)
    k = cadr(fields_reg)
    k_reg = k
    value_reg = n + value_reg
    pc = apply_cont

def sum_cps():
    global value_reg, pc, k_reg, n_reg
    if n_reg == 0:
        value_reg = 0
        pc = apply_cont
    else:
          k_reg = make_cont(cont_2, n_reg, k_reg)
        n_reg = n_reg - 1
        pc = sum_cps

# top-level function
def sum(n):
    global k_reg, n_reg, pc
    k_reg = make_cont(cont_1)
    n_reg = n
    pc = sum_cps
    return trampoline()

def car(lst):
    return lst[0]

def cdr(lst):
    return lst[1:]

def cadr(lst):
    return car(cdr(lst))

def cddr(lst):
    return cdr(cdr(lst))
\end{verbatim}
}
\caption{Python register machine code}
\label{fig:pythonRM}
\end{minipage}
\end{figure}


In a similar fashion, we transform our Calysto Scheme interpreter from a
high-level recursive CPS program written in Scheme into an equivalent low-level
register machine written in Python, which does not grow Python's call
stack.\footnote{Although the register machine does not use Python's stack,
  Calysto Scheme can nevertheless generate stack-like tracebacks from
  continuations for debugging purposes if desired. See Figure~\ref{fig:fact2}
  below for an example.}  After the build process completes, the output is a
single Python file with no dependencies, which can be easily installed and
used.

There is an interesting aspect to creating a language in this manner: the
implementation can be tested at each stage of the transformation process to
ensure correctness. That is, the same suite of Scheme test programs can be run
independently with the CPS, Data Structures, Register Machine, and Python
implementations. In essence, the initial CPS definition serves as both a
high-level language specification and an executable implementation of the
language, from which the other three implementations are subsequently
derived. This allows us to test each stage separately to catch bugs in the CPS
specification and the transformation process itself.

The current version of Calysto Scheme requires an existing Scheme
implementation to carry out the transformations from CPS to Data Structures,
and from Data Structures to Register Machine.  The final transformation step
from Scheme to Python is carried out by a Python program.  We used Petite Chez
Scheme in early development, and switched to Chez Scheme when it was made open
source. In principle, it would be possible to make Calysto Scheme be
self-hosting (\emph{i.e.}, Calysto Scheme could carry out the transformations
itself). However, our transformation program currently relies on Chez Scheme's
\texttt{syntax-rules} macro definition facility, which differs somewhat from
the version of \texttt{define-syntax} implemented in Calysto Scheme, as
explained in the next section.

\pagebreak

\section{Syntactic Extension}

\noindent
Calysto Scheme supports syntactic extension through its own version of
\texttt{define-syntax}, which can define simple macros using standard list
notation in conjunction with unification pattern matching variables beginning
with the \texttt{?} character.  For example, consider the following macro
expansion rules for the logical expressions \texttt{and} and \texttt{or}:\\

\begin{minipage}{\textwidth}
\texttt{(and $\mathit{exp}$)} $~\rightarrow~$ $\mathit{exp}$\\
\texttt{(and $\mathit{exp}_1~\mathit{exp}_2~\mathit{exp}_3~\ldots$)} $~\rightarrow~$
\texttt{(if $\mathit{exp}_1$ (and $\mathit{exp}_2~\mathit{exp}_3~\ldots$) \#f)}\\

\texttt{(or $\mathit{exp}$)} $~\rightarrow~$ $\mathit{exp}$\\
\texttt{(or $\mathit{exp}_1~\mathit{exp}_2~\mathit{exp}_3~\ldots$)} $~\rightarrow~$
\texttt{(if $\mathit{exp}_1$ \#t (or $\mathit{exp}_2~\mathit{exp}_3~\ldots$))}\\
\end{minipage}

\noindent
Although \texttt{and} and \texttt{or} are already available as special forms in
Calysto Scheme, in principle they could be implemented with the following
recursive macro definitions:\\

\begin{minipage}{\textwidth}
\begin{verbatim}

(define-syntax and
  [(and ?exp) ?exp]
  [(and ?first-exp . ?other-exps) (if ?first-exp (and . ?other-exps) #f)])

\end{verbatim}
\end{minipage}

\begin{minipage}{\textwidth}
\begin{verbatim}

(define-syntax or
  [(or ?exp) ?exp]
  [(or ?first-exp . ?other-exps) (if ?first-exp #t (or . ?other-exps))])

\end{verbatim}
\end{minipage}

\noindent
For example, the input expression \texttt{(or a b c d)} would match the pattern
of the second clause above, with the variable \texttt{?first-exp} bound to
\texttt{a} and the variable \texttt{?other-exps} bound to the list
\texttt{(b~c~d)}. The corresponding template subexpression
\texttt{(or~.~?other-exps)} would then get instantiated as \texttt{(or~b~c~d)},
and recursively expanded further. This type of pattern matching facility is
similar to Chez Scheme's \texttt{syntax-rules} form, except that instead of
using ellipses to match sequences of expressions, Calysto Scheme's
\texttt{define-syntax} uses improper lists.

Compared to \texttt{syntax-rules}, however, this approach has some limitations
and cannot describe certain transforms that can be expressed succinctly and
naturally using ellipses.  For example, to transform the pattern
\texttt{(let~((i~e)~...)~b)} into \texttt{((lambda~(i~...)~b)~e~...)} with
\texttt{syntax-rules}, ellipses are used to separate the variable bindings into
their constituent identifiers and value expressions, but this cannot be done as
easily with Calysto Scheme's unification pattern matching
notation.\footnote{The \texttt{let} macro can be defined in Calysto Scheme
  using an auxiliary ``helper'' definition, as shown below:
\begin{verbatim}
(define-syntax let
  [(let ?bindings . ?bodies) (let-helper ?bindings () () . ?bodies)])

(define-syntax let-helper
  [(let-helper () ?ids ?exps . ?bodies) ((lambda ?ids . ?bodies) . ?exps)]
  [(let-helper ((?i ?e) . ?other-bindings) ?ids ?exps . ?bodies)
   (let-helper ?other-bindings (?i . ?ids) (?e . ?exps) . ?bodies)])
\end{verbatim}
}  Furthermore, macros in Calysto Scheme are not hygienic.

Although Calysto's \texttt{define-syntax} form is not as convenient or
powerful as \texttt{syntax-rules} or \texttt{syntax-case}, it does provide
the ability to easily define a wide range of syntactic transforms via pattern
matching.  Implementing a sophisticated syntactic extension system, like that
provided in Chez Scheme, was not a central goal of the Calysto project, but we
nevertheless felt that it was important to include at least basic support for
user-defined macros in Calysto Scheme.

\section{Example: Nondeterministic Backtracking}

\noindent
In their classic text \emph{Structure and Interpretation of Computer Programs}
(SICP) \cite{SICP}, Abelson and Sussman introduce the nondeterministic ``amb''
operator for automatic backtracking.  We have incorporated this operator into
Calysto Scheme as the special form
\texttt{(choose~$\mathit{arg}_1~\mathit{arg}_2~\ldots~\mathit{arg}_n$)}, which
nondeterministically chooses one of its arguments to evaluate and returns the
resulting value.  From there, the computation proceeds normally, unless
\texttt{(choose)} is subsequently invoked with no arguments, at which point the
computation ``fails'' and immediately jumps back to the previous
\texttt{choose} expression, whereby a different argument is chosen to evaluate
next, and the computation restarts from that point with the new value.  Many
constraint-satisfaction problems can be elegantly solved using \texttt{choose}.

As an extended example illustrating both \texttt{choose} and
\texttt{define-syntax}, suppose we wish to write a program to determine how to
color a map of (a portion of) Western Europe using four distinct colors.  We
first define a function to nondeterministically return one of four possible
colors:

{\small
\begin{verbatim}
(define choose-color
  (lambda ()
    (choose 'red 'yellow 'blue 'white)))
\end{verbatim}
}

\noindent
The \texttt{color-europe} program begins by nondeterministically assigning a
color to each country on the map:

{\small
\begin{verbatim}
(define color-europe
  (lambda ()
    (let ([portugal (choose-color)]
          [spain (choose-color)]
          [france (choose-color)]
          [belgium (choose-color)]
          [germany (choose-color)]
          [luxembourg (choose-color)]
          [italy (choose-color)]
          [switzerland (choose-color)])
        ...)))
\end{verbatim}
}

\noindent
Next, we must apply the following constraint to each country: its chosen color
must be different from that of all of its adjacent neighbors.  For example,
since Luxembourg is bordered by France, Belgium, and Germany, we could express
its color constraint as follows:

{\small
\begin{verbatim}
(require (not (member luxembourg (list france belgium germany))))
\end{verbatim}
}

\noindent
The \texttt{require} function is a built-in Calysto Scheme primitive similar to
an assertion statement, which takes a boolean value as input and invokes
\texttt{(choose)} if the input is false in order to force the program to
backtrack to the most recent choice point, instead of raising an exception.
However, a more elegant approach might be to define a new syntactic form called
\texttt{color} that expands to the above code but expresses the constraint in a
more readable way:

{\small
\begin{verbatim}
(color luxembourg different from france belgium germany)
\end{verbatim}
}

\noindent
The Calysto Scheme macro definition for \texttt{color} is given below, along
with the complete program to determine a consistent map-coloring:

{\small
\begin{verbatim}
(define-syntax color
  [(color ?country different from . ?neighbors)
   (require (not (member ?country (list . ?neighbors))))])
\end{verbatim}
}

\noindent
\begin{minipage}{\textwidth}
{\small
\begin{verbatim}
(define color-europe
  (lambda ()
    (let ([portugal (choose-color)]
          [spain (choose-color)]
          [france (choose-color)]
          [belgium (choose-color)]
          [germany (choose-color)]
          [luxembourg (choose-color)]
          [italy (choose-color)]
          [switzerland (choose-color)])
      ;; apply the constraints
      (color portugal different from spain)
      (color spain different from france portugal)
      (color france different from spain italy switzerland belgium germany luxembourg)
      (color belgium different from france luxembourg germany)
      (color germany different from france switzerland belgium luxembourg)
      (color luxembourg different from france belgium germany)
      (color italy different from france switzerland)
      (color switzerland different from france italy germany)
      ;; return a coloring that satisfies the constraints
      (list (list 'portugal portugal)
            (list 'spain spain)
            (list 'france france)
            (list 'belgium belgium)
            (list 'germany germany)
            (list 'luxembourg luxembourg)
            (list 'italy italy)
            (list 'switzerland switzerland)))))

\end{verbatim}
}
\end{minipage}

\noindent
Calling \texttt{(color-europe)} returns the solution \texttt{((portugal red)
(spain yellow) (france red) (belgium yellow) (germany blue) (luxembourg
white) (italy yellow) (switzerland white))}. However, this is not the only
valid solution; many other color combinations will work.  We can force the
program to backtrack to find another combination that satisfies the
constraints, simply by calling \texttt{(choose)} as many times as we like,
until all valid choices have been returned:

{\small
\begin{verbatim}
==> (choose)
((portugal red) (spain yellow) (france red) (belgium yellow) (germany blue)
 (luxembourg white) (italy blue) (switzerland yellow))
==> (choose)
((portugal red) (spain yellow) (france red) (belgium yellow) (germany blue)
 (luxembourg white) (italy blue) (switzerland white))
\end{verbatim}
}

\noindent
and so on.  Eventually, after all possible combinations of choices satisfying
the constraints have been returned, any subsequent calls to \texttt{(choose)}
will return the string ``no more choices'', until a new \texttt{choose}
expression is executed.

\section{Python / Scheme Interoperation}

There are a number of ways that Calysto Scheme and Python can interoperate. For
example, Python code can be directly evaluated from within Scheme. Scheme
functions can be defined in an environment shared with Python, and then called
by Python programs. And Python functions and libraries can be imported directly
into Scheme and called from within Scheme programs.

Whereas in Scheme the function \texttt{eval} can be used to evaluate all types
of expressions, Python makes a distinction between evaluating expressions and
executing statements. Thus in Calysto Scheme, the function \texttt{python-eval} is
available to evaluate strings representing Python expressions, and
\texttt{python-exec} is used to execute Python statements.\\

{\small
\noindent
\texttt{(python-eval "1 + 2")}\\
$\rightarrow$ 3

\noindent
\begin{verbatim}
(python-exec
"
def multiply(a, b):
    return a * b
")
\end{verbatim}
\noindent\texttt{(python-eval "multiply(2, 3)")}\\
$\rightarrow$ 6\\
}

\noindent
The special form \texttt{func} turns a Scheme procedure into a Python function,
and \texttt{define!} puts it into the shared environment with Python:\\

{\small
\noindent
\texttt{(define! square (func (lambda (n) (* n n))))}\\
\texttt{(python-eval "square(3)")}\\
$\rightarrow$ 9\\
}

\noindent
As a simple illustration of Calysto Scheme's ability to leverage the power of
Python libraries, consider the following example, in which the elements of a
nested Scheme list are summed by converting it to a NumPy array and then
applying the NumPy \texttt{sum} function, instead of first flattening the list
or recursively summing the sublists:\\

\noindent
\begin{minipage}{\textwidth}
{\small
\begin{verbatim}
(import "numpy")
(let ([matrix3d '(((10 20 30) (40 50 60))
                  ((70 80 90) (100 110 120))
                  ((130 140 150) (160 170 180))
                  ((190 200 210) (220 230 240)))])
  (numpy.sum (numpy.array matrix3d)))
\end{verbatim}
$\rightarrow$ 3000\\
}
\end{minipage}

\noindent
Plotting data with Python's Matplotlib library is likewise a simple matter in
Calysto Scheme:\\

{\small
\begin{verbatim}
(import-as "matplotlib.pyplot" "plt")
(plt.plot '(1 2 3 4 5) '(1 4 9 16 25))
(plt.show)

\end{verbatim}
}

\noindent
Python dictionaries can also be created and manipulated from within Calysto
Scheme using a special syntax, as shown in the example interaction below:\\

\noindent
\begin{minipage}{\textwidth}
{\small
\begin{verbatim}
==> (define d (dict '((apple : red) (banana : yellow) (lime : green))))
==> d
{'apple': red, 'banana': yellow, 'lime': green}
==> (get-item d 'apple)
red
==> (set-item! d 'apple 77)
==> d
{'apple': 77, 'banana': yellow, 'lime': green}
\end{verbatim}
}
\end{minipage}

\noindent
A more complex example that shows the power of combining these features is
shown in Figure~\ref{fig:tensorflow}. This Scheme code sets up a series of
machine learning experiments using the NumPy and TensorFlow libraries imported
from Python, along with a third-party library called \texttt{comet\_ml} that
provides tools for managing experiments.  The \texttt{choose} operator is used
to select a particular combination of ``hyperparameters'' for an experiment,
and a neural network is then defined using TensorFlow functions.  Many
TensorFlow functions are configured via keyword parameters in Python, which
here we provide as Calysto Scheme dictionaries.  After an experiment using a
particular set of hyperparameters has run to completion, we can simply invoke
\texttt{(choose)} to pick a new combination of hyperparameters and rerun the
experiment.


\begin{figure}[h]
\begin{minipage}{0.8\textwidth}
{\scriptsize
\begin{verbatim}
(import-as "tensorflow" "tf")
(import-as "numpy" "np")
(import "comet_ml")

;; load the MNIST dataset using the Keras load_data function
(define dataset (tf.keras.datasets.mnist.load_data))

;; prepare the dataset
(define x_train (/ (get-item (get-item dataset 0) 0) 255.0))
(define y_train (get-item (get-item dataset 0) 1))
(define x_test (/ (get-item (get-item dataset 1) 0) 255.0))
(define y_test (get-item (get-item dataset 1) 1))

(define loss_fn (tf.keras.losses.SparseCategoricalCrossentropy (dict '((from_logits : #t)))))

(let* ([optimizer (choose "adam" "rmsprop" "sgd")]
       [dropout_rate (choose 0.0 0.1 0.2 0.4)]
       [activation (choose "relu" "sigmoid")]
       [hidden_layer_size (choose 10 20 30)]
       [options (dict `((optimizer : ,optimizer)
                        (loss : ,loss_fn)
                        (metrics : ,(vector "accuracy"))))]
       [epochs 5]
       [experiment (comet_ml.Experiment (dict '((project_name : "calysto-scheme"))))]
       [model (tf.keras.models.Sequential 
                (vector
                 (tf.keras.layers.Flatten (dict '((input_shape : (28 28)))))
                     (tf.keras.layers.Dense hidden_layer_size (dict `((activation : ,activation))))
                 (tf.keras.layers.Dropout dropout_rate)
                 (tf.keras.layers.Dense 10)))])
  (print experiment.url)
  (model.compile options)
  (model.summary)
  (experiment.log_parameters (dict `((optimizer : ,optimizer)
                                     (dropout_rate : ,dropout_rate)
                                     (activation : ,activation)
                                     (hidden_layer_size : ,hidden_layer_size)
                                     (epochs : ,epochs)))
                             (dict))
  (experiment.set_model_graph model)
  (let ([history (model.fit x_train y_train (dict `((epochs : ,epochs))))]
        [step 0])
    (map (lambda (key)
           (set! step 0)
           (map (lambda (v)
                  (experiment.log_metric key v step)
                  (set! step (+ step 1)))
                (get-item history.history key)))
         history.history))
  (experiment.end))
\end{verbatim}
}
\end{minipage}
\caption{Calysto Scheme code for running TensorFlow experiments.}
\label{fig:tensorflow}
\end{figure}


\section{Calysto Scheme in Jupyter}

The Jupyter Project is a language-independent client-server system for
running code using a web browser as the IDE
\cite{Kluyver2016jupyter}. Initially, Jupyter was limited to Python
(and was originally called IPython \cite{PER-GRA:2007}). However, the
developers realized that their system could be used for any language,
and the Jupyter Project was born. Today, Jupyter is one of the
most-used tools in data science. Jupyter Notebooks allow the mixing of
code, text, mathematics, and visualizations within a single executable
document.

In many ways, the goals of Jupyter were similar to the original goals of our
Calico project: language independence with a common UI. In fact, when Jupyter
was announced, we realized that this made more sense than our C\#-based system,
and we began migration of our project to the Jupyter framework. Calysto Scheme,
in fact, gets its name from Callisto, the second largest moon of the planet
Jupiter.

Running Calysto Scheme in a Jupyter environment (such as notebook, jupyterlab,
console, or qtconsole) provides the following additional features:

\begin{itemize}
\item TAB completions of Scheme functions and variable names
\item Ability to directly display rich media such as images
\item Access to so-called magics (\texttt{\%} meta-commands)
\item Ability to easily run shell commands using the ``\texttt{!} command'' format
\item \LaTeX-style equations and variables
\end{itemize}

\begin{figure}[h]
\begin{minipage}{0.47\textwidth}
  \includegraphics[width=\textwidth]{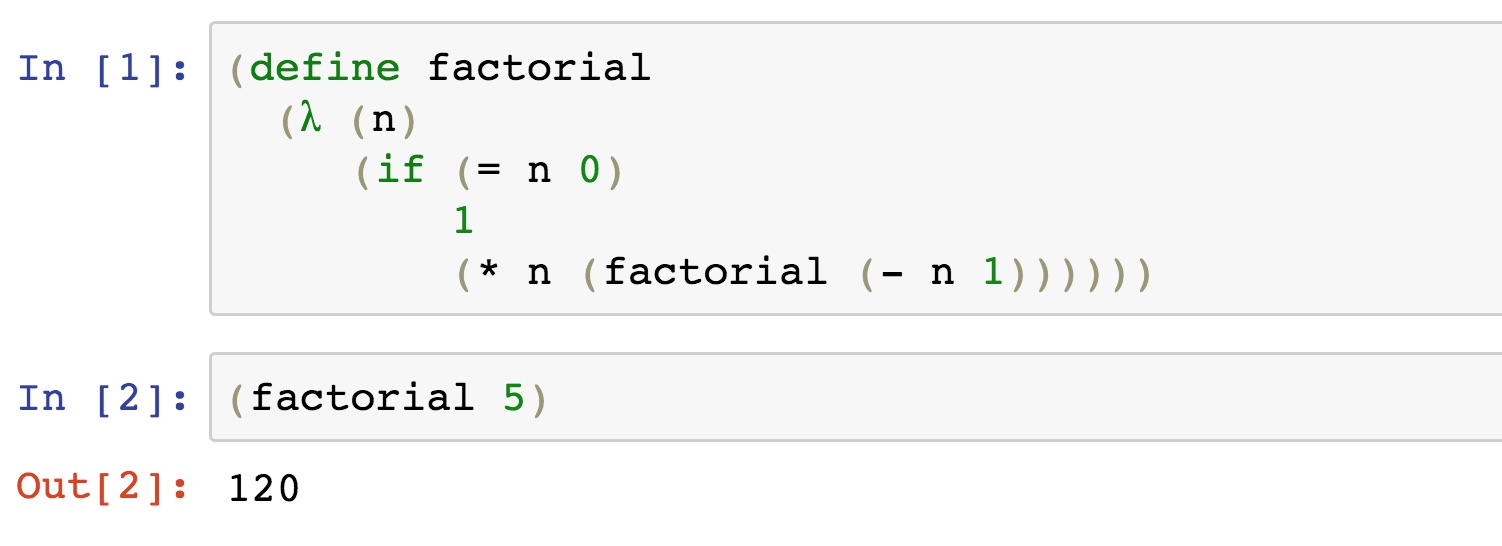}
  \caption{Calysto Scheme running in a Jupyter Notebook.}
  \label{fig:fact1}
  \Description{Calysto Scheme code for the factorial function.}
\end{minipage}
\hspace{0.2in}
\begin{minipage}{0.47\textwidth}
  \includegraphics[width=\textwidth]{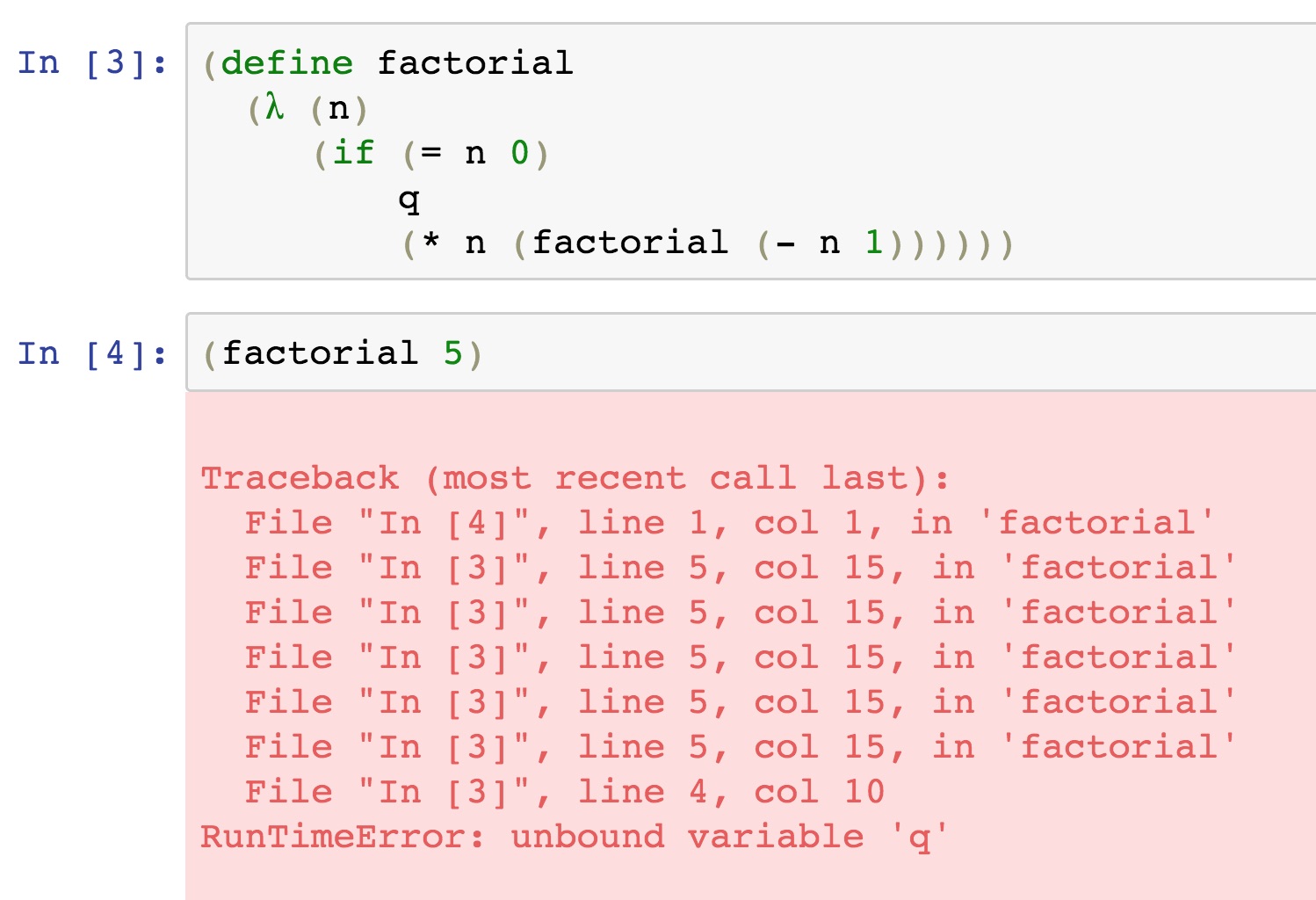}
  \caption{Example of a traceback in Calysto Scheme.}
  \label{fig:fact2}
  \Description{An error message in the Calysto Scheme programming language.}
\end{minipage}
\end{figure}

For example, in Calysto Scheme running in Jupyter, one can create a lambda
function with the Greek symbol $\lambda$ in place of the \texttt{lambda}
keyword by typing \textbf{\textbackslash lambda} and pressing [TAB] (see
Figure~\ref{fig:fact1}). Building on top of the Jupyter system makes Calysto
Scheme easy to use, and brings it into a modern language environment.


Stack-like tracebacks are also available to help with debugging. For example,
in Figure~\ref{fig:fact2} the definition of the factorial function contains a
typo, with \texttt{q} instead of 1 for the base case, which generates an
unbound variable error. The traceback shows the sequence of recursive calls
that were made before the error occurred. In many Scheme implementations, since
control is not managed with a stack, it is not possible to generate such a
stack trace. However, in Calysto Scheme stack tracing can be enabled or
disabled via the parameter \texttt{use-stack-trace}. For example, calling
\texttt{(use-stack-trace~\#f)} turns off stack tracing.

Many Jupyter language kernels (including IPython's) support a set of
meta-commands called ``magics''. These meta-commands allow the user to easily
perform a variety of functions such as running operating system commands,
creating files, timing execution, and many more. We wanted the Calysto Scheme
Jupyter kernel also to have easy-to-use meta-commands. However, rather than
building those directly into the Calysto Scheme kernel, we decided to put them
in a separate abstraction layer. Thus, we collaborated with others to create a
project called ``Metakernel'' \cite{Metakernel}. Metakernel supports many
meta-commands, including many designed for educational use, and does so in a
language-independent manner. Metakernel's GitHub page currently lists over 700
different projects that have incorporated it in one way or another. In fact,
Metakernel may be the most widely used contribution of the Calysto Scheme
Project to date.

The full list of Metakernel's magics can be found in
\cite{MetakernelMagics}. All of these extend Calysto Scheme to make it more
useful in Jupyter environments. However, one meta-command that deserves special
mention in this context is the \texttt{\%parallel} magic. If a parallel
computing cluster is available, this magic turns Calysto Scheme into a
parallelized interpreter. For example, the following command initializes the
cluster for running Calysto Scheme:

{\small
\begin{verbatim}
%parallel calysto_scheme CalystoScheme
\end{verbatim}
}

\begin{figure}[h]
  \centering
  \includegraphics[width=0.65\textwidth]{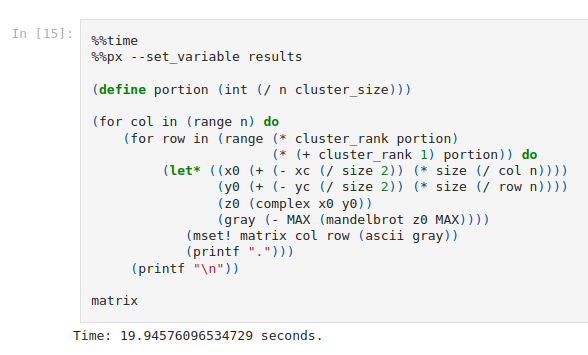}
  \caption{Parallel processing code for generating the Mandelbrot set in
    Calysto Scheme, from \cite{ParallelProcessingWithMetakernel}.}
  \label{fig:parallelized-scheme}
  \Description{Parallelized Calysto Scheme.}
\end{figure}

After initializing the cluster, individual Scheme expressions or entire code
cells can be executed in parallel using the \texttt{\%px}
command. Figure~\ref{fig:parallelized-scheme} shows an excerpt of Calysto
Scheme code for computing the Mandelbrot set in parallel. The full example
can be found in \cite{ParallelProcessingWithMetakernel}.


It should be noted that some computer science educators find the use of Jupyter
Notebooks to be problematic \cite{Johnson2020}. The main objection is that it
is easy for a student to create hidden state that may lead to unexpected and
non-intuitive results. For example, one can edit and rerun cells that have
already been executed, introducing new dependencies that may be difficult to
reproduce. On the other hand, we have also heard educators lament that
``Jupyter is \emph{too} convenient'' for the student.

In spite of these warnings, Jupyter Notebooks are popular with data scientists
and others who tell stories with code, visualizations, and text. We believe
this can make an excellent environment for teaching and learning. One of us
(Blank) taught all of his classes for four years using nothing but
Jupyter. That included courses on Data Structures (Java), Cognitive Science
(Python), Assembly Language, Introduction to Computer Science (Processing),
Programming Languages (Scheme), and Creative Writing (English). This was
accomplished by hosting a local JupyterHub instance\footnote{Incidentally,
  this was the very first publicly-hosted JupyterHub instance.} at Bryn Mawr
College, so that students did not need to install anything and could perform
all of their work via the web browser. For more on these ideas, see
\cite{Calico2}.

\begin{figure}[h]
  \centering
  \includegraphics[width=0.4\textwidth]{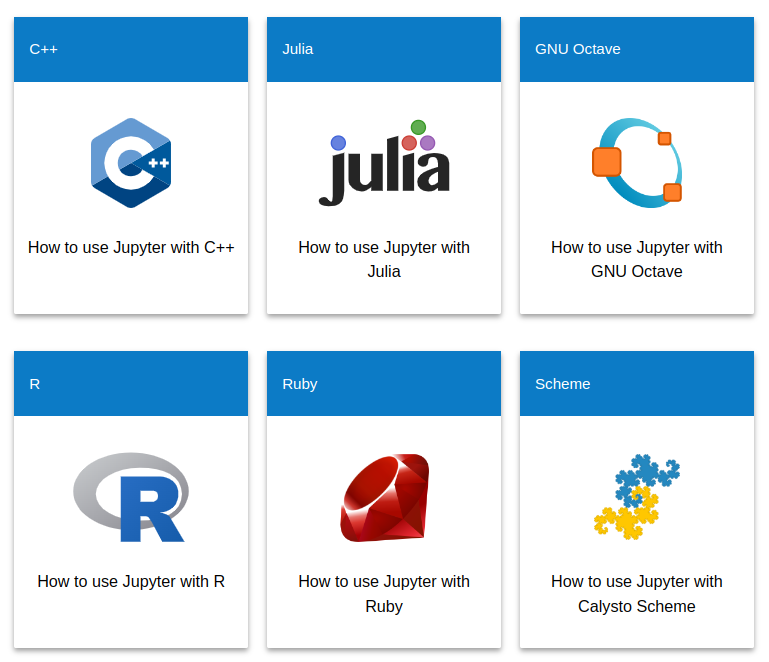}
  \caption{Calysto Scheme is available online via the Jupyter Project's ``Try Jupyter'' page.}
  \label{fig:try-jupyter}
  \Description{Calysto Scheme, online.}
\end{figure}

Judging from the Calysto Scheme issues reported on GitHub, most people discover
Calysto Scheme through the Jupyter Project's ``Try Jupyter'' page
\cite{TryJupyter}, where Calysto Scheme is featured along with kernels for C++,
Julia, GNU Octave, R, and Ruby. See Figure~\ref{fig:try-jupyter}. Many people
apparently also use Calysto Scheme for working through SICP exercises.


\section{Scheme in Python, Python in Scheme}

An approach commonly taken in teaching a course on Programming Languages
is to implement an interpreter for a subset of Scheme in another language, such
as Python. This is a relatively straightforward process, which is considerably
simplified by Scheme's easy-to-parse syntax.  For example, the following code
shows part of an interpreter written in Python with the functions
\texttt{evaluator} and \texttt{apply\_operator} for evaluating Scheme
expressions and applying Scheme functions, respectively:\\

\begin{minipage}{\textwidth}
{\small
\begin{verbatim}
# Scheme-in-Python interpreter

# define parser, reader, tokenizer, and utilities (Map, car, cdr, etc.)

def evaluator(expr):
    if car(expr) == "literal":
        return cadr(expr)
    elif car(expr) == "application":
        return apply_operator(evaluator(cadr(expr)),
                              Map(evaluator, caddr(expr)))
    else:
        raise Exception("Invalid AST: %s" % expr)

def apply_operator(op, operands):
    if op == "+":
        return sum(operands)
    else:
        raise Exception("Unknown operator: %s" % op)

\end{verbatim}
\texttt{evaluator(parser(reader(tokenizer("(+ 1 2)"))))}\\
$\rightarrow$ 3\\
}
\end{minipage}

\noindent
Conversely, one could also teach programming language principles by
implementing Python in Scheme. However, this would be much more challenging to
attempt in a single semester course, due to the complexity of Python's
syntax. However, if it were possible to outsource the parsing of Python syntax
into Abstract Syntax Tree (AST) structures, then one could commence with
building a Python interpreter written in Scheme that operates directly on
Python ASTs. Because Calysto Scheme can call Python libraries, this becomes a
simple task, as the \texttt{ast} Python library has the ability to take strings
of Python code and turn them into ASTs \cite{PythonInScheme}.

Mirroring the above Python functions for interpreting Scheme code, one can
easily construct similar functions written in Scheme for interpreting Python
ASTs:\\

\begin{minipage}{\textwidth}
{\small
\begin{verbatim}
;; Python-in-Scheme interpreter

(import "ast")

(define evaluator
  (lambda (ast_expr)
   (cond
     [(isinstance ast_expr ast.Module)
      (evaluator (get-item ast_expr.body 0))]
     [(isinstance ast_expr ast.Num)
      ast_expr.n]
     [(isinstance ast_expr ast.Expr)
      (evaluator ast_expr.value)]
     [(isinstance ast_expr ast.BinOp)
      (apply-operator ast_expr.op
         (evaluator ast_expr.left)
         (evaluator ast_expr.right))]
     [else (error 'evaluator (format "Unknown ast: ~s" ast_expr))])))

(define apply-operator
  (lambda (op v1 v2)
    (cond
      [(isinstance op ast.Add) (+ v1 v2)]
      [else (error 'apply-operator (format "Invalid operator: ~s" op))])))

\end{verbatim}
\texttt{(evaluator (ast.parse "1 + 2"))}\\
$\rightarrow$ 3\\
}
\end{minipage}

\noindent
The main differences between the Scheme interpreter in Python, and the Python
interpreter in Scheme arise from the different ways in which they represent
abstract syntax.  For example, the Scheme expression \texttt{(+~1~2)} is
treated as a function application AST by the Scheme-in-Python interpreter,
whereas the Python expression \texttt{"1~+~2"} is treated as a binary operator
AST by the Python-in-Scheme interpreter.  One could easily change the Scheme
abstract syntax structures to more closely mimic the Python AST structures if
desired, in order to make the two interpreters more symmetric.

\section{Summary}

In this paper we have presented Calysto Scheme, a full-featured implementation
of the Scheme language written in Scheme and transpiled into Python, with
support for easy interoperation with many Python libraries, as well as Jupyter
Notebooks. This system was originally designed primarily with CS education in
mind, and has found many novel uses, including as an effective platform for
teaching programming language concepts through the hands-on development of
Scheme interpreters in Python, and, conversely, Python interpreters in
Scheme. Although our focus has been mainly on pedagogy, we are gratified that
others have found Calysto Scheme to be useful in more general contexts as well.

\begin{acks}
We would like to express our gratitude to Professor Dan Friedman and the
Indiana University programming languages community, for instilling in us a deep
and abiding appreciation for the programming language concepts and techniques
that inspired the creation of Calysto Scheme.  We would also like to thank
Steven Silvester for his constant and continued management of the Calysto
repository. Thanks also to Jason Hemann for many helpful comments on an
earlier draft of this paper, and to the anonymous reviewers for providing
additional useful suggestions.
\end{acks}

\bibliographystyle{ACM-Reference-Format}
\bibliography{scheme2023}

\end{document}